\documentclass[aip,reprint,amsmath,amssymb,twocolumn,twoside,nofootinbib]{revtex4-1}

\usepackage{listings} 

\usepackage{bbold}
\usepackage{amsmath,amssymb,latexsym}    
\usepackage{wasysym}
\usepackage{amsfonts}
\usepackage{siunitx}
\usepackage{blindtext}
\usepackage{units}

\usepackage{graphicx}
\usepackage{dcolumn}
\usepackage{bm}
\usepackage{gensymb}

\usepackage[symbol*]{footmisc}
\DefineFNsymbolsTM{myfnsymbols}{
  \textasteriskcentered *
  \textdagger    \dagger
  \textdaggerdbl \ddagger
  \textsection   \mathsection
  \textbardbl    \|%
  \textparagraph \mathparagraph
}%
\setfnsymbol{myfnsymbols}

\newcommand{\dg}[1]{\ensuremath{#1^\circ}}
\newcommand{\wavenumber}[1]{\ensuremath{#1~\text{cm}^{-1}}}
\newcommand{\mumetr}[1]{\ensuremath{\SI{#1}{\micro\meter}}}
\newcommand{\nmetr}[1]{\ensuremath{#1~\text{nm}}}

\newcommand{\kvec}{\ensuremath{\vec{k}}}
\newcommand{\etal}[1]{#1 \textit{et al}.}
\newcommand{\AMa}[1]{\ensuremath{\bold{A}_{#1}}}
\newcommand{\PMa}[1]{\ensuremath{\bold{P}_{#1}}}
\newcommand{\TMa}[1]{\ensuremath{\bold{T}_{#1}}}
\newcommand{\LMa}[1]{\ensuremath{\bold{L}_{#1}}}
\newcommand{\YTME}[1]{\ensuremath{\Gamma_{#1}^*}}
\newcommand{\epsiTens}{\ensuremath{\bar{\varepsilon}}}
\newcommand{\incAngle}{\ensuremath{\theta}}
\newcommand{\rotMatrix}{\ensuremath{\Omega}}
\newcommand{\epsi}{\ensuremath{\varepsilon}}
\newcommand{\epsiPerp}{\ensuremath{\epsi_{\perp}}}
\newcommand{\epsiPar}{\ensuremath{\epsi_{\parallel}}}
\newcommand{\qij}{\ensuremath{q_{ij}}}
\newcommand{\parDer}[1]{\ensuremath{\frac{\partial}{\partial #1}}}

\newcommand{\colvec}[1]{\ensuremath{\begin{pmatrix}#1\end{pmatrix}}}
\newcommand{\rhoOneT}{\ensuremath{\bar{\rho}_1}}
\newcommand{\rhoTwoT}{\ensuremath{\bar{\rho}_2}}
\newcommand{\muT}{\ensuremath{\bar{\mu}}}
\newcommand{\Del}{\ensuremath{\bold{\Delta}}}
\newcommand{\fracM}[2]{\ensuremath{\left( #1 \right) / #2}}
\newcommand{\abs}[1]{\ensuremath{\lvert #1 \rvert}} 
\newcommand{\Hvec}{\ensuremath{\vec{H}}}

\newcommand{\Evec}{\ensuremath{\vec{E}}}

\newcommand{\ga}[1]{\ensuremath{\gamma_{#1}}}
\newcommand{\FullTMa}[1]{\ensuremath{\bold{\Gamma}_{#1}}}
\newcommand{\YehTMa}[1]{\ensuremath{\bold{\Gamma}_{#1}^*}}
\newcommand{\twotwoMatrix}[4]{\ensuremath{\begin{pmatrix}
#1 & #2 \\ 
#3 & #4 \end{pmatrix}}}
\newcommand{\threethreeMatrix}[9]{\ensuremath{\begin{pmatrix}
#1 & #2 & #3 \\
#4 & #5 & #6 \\ 
#7 & #8 & #9 \end{pmatrix}}}
\newcommand{\fourfourMatrix}[4]{\ensuremath{\begin{pmatrix}
#1 \\
#2 \\ 
#3 \\
#4 \end{pmatrix}}}

\newcommand{\sixsixMatrix}[6]{\ensuremath{\begin{pmatrix}
#1 \\
#2 \\ 
#3 \\
#4 \\
#5 \\
#6 \end{pmatrix}}}
\newcommand{\twopartdef}[4]
{
	\left\{ \!\!
		\begin{array}{ll}
			#1, &\!\! \mbox{} #2 \\
			#3, &\!\! \mbox{} #4
		\end{array}
	\right.
}

\begin{document}

\preprint{APS/123-QED}

\title{Generalized $\bf{4\times 4}$ Matrix Formalism for Light Propagation in Anisotropic Stratified Media: Study of Surface Phonon Polaritons in Polar Dielectric Heterostructures}

\author{Nikolai Christian Passler}
\altaffiliation{Fritz-Haber-Institut der Max-Planck-Gesellschaft, Faradayweg 4-6,14195 Berlin, Germany}
\email{passler@fhi-berlin.mpg.de}
\author{Alexander Paarmann}
\altaffiliation{Fritz-Haber-Institut der Max-Planck-Gesellschaft, Faradayweg 4-6,14195 Berlin, Germany}

\date{\today}



\begin{abstract}
We present a generalized $4 \times 4$ matrix formalism for the description of light propagation in birefringent stratified media. In contrast to previous work, our algorithm is capable of treating arbitrarily anisotropic or isotropic, absorbing or non-absorbing materials and is free of discontinous solutions. We calculate the reflection and transmission coefficients and derive equations for the electric field distribution for any number of layers. The algorithm is easily comprehensible and can be straight forwardly implemented in a computer program. To demonstrate the capabilities of the approach, we calculate the reflectivities, electric field distributions, and dispersion curves for surface phonon polaritons excited in the Otto geometry for selected model systems, where we observe several distinct phenomena ranging from critical coupling to mode splitting, and surface phonon polaritons in hyperbolic media. 
\end{abstract}

\maketitle


\section{Introduction}

Light-matter interaction in complex hybrid nanostructures has become a central problem of modern nanophotonics \cite{Maier2007}. Specifically thin films, periodic layered media or other stratified systems have proven to exhibit extensive functionality, for instance in anti-reflection coatings \cite{Hiller2002,Xi2007}, transistors \cite{Nomura2003,Street2009}, thin-film photovoltaics \cite{Chopra2004,Shin2011}, or sensing \cite{Patel2003,Yoo2005,Gong2006}.  However, the angle of incidence-dependent calculation of light propagation in a general multilayer structure rapidly becomes cumbersome, especially if optical anisotropy and absorption are taken into account. In order to solve this problem, many authors have presented $4\times 4$ matrix approaches \cite{Berreman1972,Li1988,Xu2000,Lin-Chung1984,Yeh1979,Zhang2015, OBrien2013} based on Maxwell's equations. However, many of these methods consider only special cases of the dielectric tensor \cite{Berreman1972, Yeh1979, Lin-Chung1984,OBrien2013} or lead to discontinuous solutions \cite{Xu2000, Zhang2015}.

These problems become particularly critical in the mid- and far-infrared Reststrahlen spectral region of polar dielectrics \cite{Adachi1999}, where surface phonon polaritons (SPhPs) can be excited \cite{Huber2005,Falge1973,Neuner2009,Passler2017}. Here, the highly dispersive and often strongly anisotropic behavior of the dielectric function precludes the use of formalisms that are restricted to special cases. However, it is exactly in these materials and atomic-scale heterostructures thereof \cite{Dai2014,Caldwell2015a,Caldwell2016,Woessner2015}, that SPhPs have very recently been demonstrated to enable many novel phenomena such as hyperbolic superlensing \cite{Taubner2006,Li2015} and negative refraction \cite{RodriguesdaSilva2010,Yoxall2015}. For instance, hexagonal boron nitride as one of the key components of van der Waals heterostructures \cite{Geim2013}, displays many interesting nanophotonic properties due to its naturally hyperbolic character \cite{Yoxall2015, Caldwell2014,Jacob2014, Li2015, Dai2014}. Therefore, a most general, robust, and easily implementable numerical formalism to analyze the optical response in arbitrarily anisotropic multilayer heterostructures is highly desirable.

In this work, we present a comprehensible $4\times 4$ matrix formalism, which can be straight-forwardly implemented in a computer program \cite{Passler2017a}. Our algorithm combines several previous approaches \cite{Berreman1972,Yeh1979,Li1988,Lin-Chung1984,Xu2000} in such a way that numerical instabilities and discontinuous solutions are prevented, to enable a robust treatment of light incident on any number of arbitrarily anisotropic or isotropic, absorbing or non-absorbing layers. In addition to reflection and transmission coefficients, we also calculate the full electric field distributions throughout the heterostructure, which is particularly useful for analysis of polariton modes and their associated local field enhancements. To demonstrate the capabilities of our algorithm, we present simulation results for several model systems where SPhPs can be excited using prism coupling in the Otto geometry \cite{Falge1973}: 6H-SiC, GaN/SiC, \mbox{SiC/GaN/SiC}, and $\alpha$-quartz, covering a number of phenomena such as critical coupling to SPhPs~\cite{Neuner2009,Passler2017}, index-sensing \cite{Neuner2010}, mode-splitting \cite{Novikova2013}, wave-guiding \cite{Zheng2016}, and polaritons in hyperbolic media \cite{RodriguesdaSilva2010,Caldwell2014}.

\section{Theory}

\subsection{Matrix Formalism}
The incident medium is assumed to be non-absorptive and isotropic, and the magnetic permeability $\mu$ is taken as a scalar. The coordinate system is defined such that the multilayer surfaces are parallel to the $x$-$y$-plane and the $z$-direction is orthogonal to the surface, pointing from the incident medium towards the substrate and being zero at the first interface between incident medium and layer $i=1$. The layers are indexed from $i=1$ to $i=N$, whereas the incident medium is $i=0$ and the substrate $i=N+1$. Each layer has an individual thickness $d_i$, and the thickness of the complete multilayer system is $D=\sum_{i=1}^{N}d_i$. Furthermore, each medium $i$ is characterized by an individual dielectric tensor $\epsiTens_i$. The incident beam is chosen to propagate in the $x$-$z$-plane with a wave vector $\kvec_i$ in layer $i$:
\begin{align}
\kvec_i=\frac{\omega}{c}(\xi,0,q_i),
\end{align}
where $\xi=\sqrt{\epsi_{inc}} \: \sin(\theta)$ is the $x$-component of $\kvec_i$ which is conserved throughout the complete multilayer system, $\epsi_{inc}$ is the isotropic dielectric constant of the incident medium, \incAngle~is the incident angle, and $q_i$ is the dimensionless $z$-component of the wave vector in layer $i$. 

For a given diagonal dielectric tensor \epsiTens~with principle dielectric constants $\epsi_x$, $\epsi_y$, and $\epsi_z$, the following transformation allows to rotate the crystal orientation into the lab frame:
\begin{align}
\epsiTens^{\:\prime}=\rotMatrix\:\epsiTens\:\rotMatrix^{-1}=\rotMatrix\: \threethreeMatrix{\epsi_x}{0}{0}{0}{\epsi_y}{0}{0}{0}{\epsi_z} \rotMatrix^{-1},
\end{align}
with the coordinate rotation matrix \rotMatrix~given by the Euler angles $\vartheta$, $\varphi$, and $\psi$ (see e.g. Eq.~2 in \cite{Yeh1979}). Furthermore, optical activity or the response to a static magnetic field can be incorporated into \epsiTens~\cite{Yariv1984}, which then becomes non-diagonal, but this will not be discussed here. 

\subsubsection{Eigenmodes in Medium $i$}
In general, each layer will have exactly four eigenmodes, i.e. four possible solutions for the propagation of an electromagnetic wave. They differ in polarization and propagation direction and have four different $z$-components of the wave vector, \qij, with $j=1,2,3,4$. In order to obtain these four solutions, we follow the approach by Berreman \cite{Berreman1972}, where the layer index $i$ is omitted for brevity.

First, the Maxwell equations are written in a $6\times 6$-matrix form, where the time derivative has already been performed, assuming monochromatic waves at frequency $\omega$:
\begin{align}
\begin{split}
\! \bold{R} \bold{G}&\equiv
\sixsixMatrix
{0 & 0 & 0 & 0 & -\parDer{z} & \parDer{y}}
{0 & 0 & 0 & \parDer{z} & 0 & -\parDer{x}}
{0 & 0 & 0 & -\parDer{y} & \parDer{x} & 0}
{0 & \parDer{z} & -\parDer{y} & 0 & 0 & 0}
{-\parDer{z} & 0 & \parDer{x} & 0 & 0 & 0}
{\parDer{y} & -\parDer{x} & 0 & 0 & 0 & 0} \!
\colvec{E_x \\ E_y \\ E_z \\ H_x \\ H_y \\H_z}\\
&=-i\omega \colvec{D_x \\ D_y \\ D_z \\ B_x \\ B_y \\ B_z} \equiv
-i\omega \bold{C}.
\end{split}
\label{eq:A_berremanMaxwell}
\end{align}
If only linear effects are considered, a relation between $\bold{G}$ and $\bold{C}$ can be formulated as follows\footnote{Please note the difference to the original work by Berremann, who instead reports $\mathbf{G=MC}$ in Eq.~(3) of Ref.~\cite{Berreman1972}.}
\begin{align}
\bold{C}=\bold{M}\bold{G} \equiv \twotwoMatrix{\epsiTens}{\rhoOneT}{\rhoTwoT}{\muT}\bold{G},
\label{eq:A_berremanLinearResponse}
\end{align}
where \muT~is the permeability tensor, and \rhoOneT~ and \rhoTwoT~are optical rotation tensors.

By combining Eq.~\ref{eq:A_berremanMaxwell} and \ref{eq:A_berremanLinearResponse}, the following spatial wave equation is obtained, where $\bold{g}$ is the spatial part of $\bold{G} = \bold{g} e^{-i\omega t}$:
\begin{align}
\bold{R} \bold{g}=-i\omega \bold{M}\bold{g}.
\label{eq:A_berremanSpatialWaveEq}
\end{align}
As shown explicitly by Berreman, the normal field projections $g_3=E_z$ and $g_6=H_z$ can be solved in terms of the other four parameters, and thus be eliminated. In short, this yields
\begin{align}
\parDer{z} \Psi=i \frac{\omega}{c} \Del \Psi,
\end{align}
where 
\begin{align}
\Psi=\colvec{E_x \\ H_y \\ E_y \\ -H_x}
\label{eq:A_reorderedFieldVec}
\end{align}
is the reordered dimensionless field vector. \Del~is exactly defined in terms of $\bold{M}$ as follows:
\begin{align}
\begin{split}
\Delta_{11}&= M_{51}+(M_{53}+\xi)a_{31}+M_{56}a_{61} \\
\Delta_{12}&= M_{55}+(M_{53}+\xi)a_{35}+M_{56}a_{65} \\
\Delta_{13}&= M_{52}+(M_{53}+\xi)a_{32}+M_{56}a_{62} \\
\Delta_{14}&=-M_{54}-(M_{53}+\xi)a_{34}-M_{56}a_{64} \\
\Delta_{21}&= M_{11}+M_{13}a_{31}+M_{16}a_{61} \\
\Delta_{22}&= M_{15}+M_{13}a_{35}+M_{16}a_{65} \\
\Delta_{23}&= M_{12}+M_{13}a_{32}+M_{16}a_{62} \\
\Delta_{24}&=-M_{14}-M_{13}a_{34}-M_{16}a_{64} \\
\Delta_{31}&=-M_{41}-M_{43}a_{31}-M_{46}a_{61} \\
\Delta_{32}&=-M_{45}-M_{43}a_{35}-M_{46}a_{65} \\
\Delta_{33}&=-M_{42}-M_{43}a_{32}-M_{46}a_{62} \\
\Delta_{34}&= M_{44}+M_{43}a_{34}+M_{46}a_{64} \\
\Delta_{41}&= M_{21}+M_{23}a_{31}+(M_{26}-\xi)a_{61} \\
\Delta_{42}&= M_{25}+M_{23}a_{35}+(M_{26}-\xi)a_{65} \\
\Delta_{43}&= M_{22}+M_{23}a_{32}+(M_{26}-\xi)a_{62} \\
\Delta_{44}&=-M_{24}-M_{23}a_{34}-(M_{26}-\xi)a_{64},
\end{split}
\end{align}
where the elements of $a_{mn}$ are given by
\begin{align}
\begin{split}
a_{3n} &= \colvec{
\fracM{M_{61}M_{36}-M_{31}M_{66}}{b} \\
 \fracM{(M_{62}-\xi)M_{36}-M_{32}M_{66}}{b} \\
 0 \\
 \fracM{M_{64}M_{36}-M_{34}M_{66}}{b} \\
 \fracM{M_{65}M_{36}-(M_{35}+\xi)M_{66}}{b} \\
 0
}  \\
a_{6n} &= \colvec{
\fracM{M_{63}M_{31}-M_{33}M_{61}}{b} \\
 \fracM{M_{63}M_{32}-M_{33}(M_{62}-\xi)}{b} \\
 0 \\
 \fracM{M_{63}M_{34}-M_{33}M_{64}}{b} \\
 \fracM{M_{63}(M_{35}+\xi)-M_{33}M_{65}}{b} \\
 0
},
\end{split}
\end{align}
and $b$ is defined by
\begin{align}
b=M_{33}M_{66}-M_{36}M_{63}.
\end{align}
Note that \Del~and $a_{mn}$, in contrast to the formulas given by Berreman, do not contain the factor $\frac{c}{\omega}$ before each $\xi$, which is because $\xi$ is chosen to be dimensionless here.

The $\bold{M}$-matrix of the corresponding material is assumed to be $z$-independent, i.e. we only consider homogeneous layers (a numerical solution if $\bold{M}$ depends on $z$ has been given by Berreman \cite{Berreman1972}). In the $z$-independent case, the four eigenvalues \qij~of $\Delta(i)$, indexed with $j$  for each layer $i$, represent the $z$-components of the wave vectors of the four eigenmodes $\Psi_{ij}$ in the material:
\begin{align}
\qij \Del(i)=\Psi_{ij} \Del(i).
\label{eq:eigensystem}
\end{align}
\\

\subsubsection{\textbf{Sorting of the Eigenmodes}}
At this point, the four solutions have to be ordered in an unambiguous manner in order to avoid unstable solutions and discontinuities. The approach presented here is based on the work of \etal{Li} \cite{Li1988}. First of all, the modes have to be separated into forward propagating (transmitted) and backward propagating (reflected) waves, which is done as follows:
\begin{align}
\begin{split}
\text{\qij~is real:}\qquad \:\:\: \qij \geq 0 \; &\longrightarrow \text{ transmitted} \\
\qij < 0 \; &\longrightarrow \text{ reflected} \\
\text{\qij~is complex:} \:\: Im(\qij) \geq 0 \; &\longrightarrow \text{ transmitted} \\
Im(\qij) < 0 \; &\longrightarrow \text{ reflected},
\end{split}
\label{eq:sorting1}
\end{align}
because real wave vectors point in the propagation direction, and complex wave vectors describe an exponentially damped wave. The transmitting waves will be labeled $q_{i1}$ and $q_{i2}$, the reflected waves $q_{i3}$ and $q_{i4}$. Each pair, however, also has to be sorted in order to ensure solutions without discontinuities. For this, the eigenvectors $\Psi_{ij}$ of Eq. \ref{eq:eigensystem} of each layer are analyzed by \cite{Li1988}
\begin{align}
C(q_{ij})=\frac{\abs{\Psi_{ij1}}^2}{\abs{\Psi_{ij1}}^2+\abs{\Psi_{ij3}}^2},
\end{align}
allowing to finally sort the four eigenvalues as follows:
\begin{align}
C(q_{i1}) > C(q_{i2}) \quad \text{and} \quad C(q_{i3}) > C(q_{i4}).
\label{eq:sorting2}
\end{align}
This means for the four solutions, that $q_{i1}$ and $q_{i3}$ describe p-polarized, and $q_{i2}$ and $q_{i4}$ s-polarized~waves, transmitted and reflected, respectively.

\subsubsection{\textbf{Transfer Matrix with Treatment of Singularities}}

In previous works \cite{Berreman1972,Yeh1979,Lin-Chung1984}, the authors assume fully anisotropic dielectric tensors, and their formalisms suffer from singularities for several special cases, i.e. if the material is isotropic, or even if the dielectric tensor has only diagonal components. These cases have in common that the four solutions of Eq. \ref{eq:eigensystem} become degenerate.  To resolve this issue, we here follow the solution presented by \etal{Xu} \cite{Xu2000}. We note, however, that the authors only considered non-optical active media with isotropic magnetic permeability, such that in the following we also set $\muT = \mu \: \bar 1$ and $\rhoOneT = \rhoTwoT = \bar 0$, with $\bar 1$~and $\bar 0$~the unity matrix and the matrix consisting just of zeros, respectively. More general solutions to avoid singularities also without these restrictions are beyond the scope of this work. 

To set up the transfer matrix according to \etal{Xu} \cite{Xu2000} using the appropriately sorted $q_{ij}$, we now write the electric field vectors $\vec{\gamma}_{ij}$ of the four eigenmodes in each layer $i$ as follows:
\begin{align}
\vec{\gamma}_{ij}=\colvec{\gamma_{ij1} \\ \gamma_{ij2} \\ \gamma_{ij3}}.
\end{align}
The values of $\gamma_{ijk}$, being free of singularities, are then given by the following formulas \cite{Xu2000}:

\begin{align}
\begin{split}
\gamma_{i11}\!&=\!\gamma_{i22}=\gamma_{i42}=-\gamma_{i31}=1, \\
\gamma_{i12}\!&= \!
\twopartdef
{0} 
{\qquad\quad\:\:\: q_{i1}\!=\!q_{i2}}
{\frac{\mu_i\epsi_{i23}(\mu_i\epsi_{i31}+\xi q_{i1})-\mu_i\epsi_{i21}(\mu_i\epsi_{i33}-\xi^2)}
{(\mu_i\epsi_{i33}-\xi^2)(\mu_i\epsi_{i22}-\xi^2-q_{i1}^2)-\mu_i^2\epsi_{i23}\epsi_{i32}}} 
{\qquad\quad\:\:\: q_{i1} \!\neq\! q_{i2}}
\\
\gamma_{i13}\!&= \!
\twopartdef
{-\frac{\mu_i\epsi_{i31}+\xi q_{i1}}{\mu_i\epsi_{i33}-\xi^2}} 
{\qquad\qquad\quad\;\; q_{i1} \!=\! q_{i2}}
{-\frac{\mu_i\epsi_{i31}+\xi q_{i1}}{\mu_i\epsi_{i33}-\xi^2} -\frac{\mu_i\epsi_{i32}}{\mu_i\epsi_{i33}-\xi^2} \gamma_{i12}} 
{\qquad\qquad\quad\;\; q_{i1} \!\neq\! q_{i2}}
\\
\gamma_{i21}\!&=\! 
\twopartdef
{0} 
{\:\:\: q_{i1} \!=\! q_{i2}}
{\frac{\mu_i\epsi_{i32}(\mu_i\epsi_{i13}+\xi q_{i2})-\mu_i\epsi_{i12}(\mu_i\epsi_{i33}-\xi^2)}
{(\mu_i\epsi_{i33}\!-\!\xi^2)\!(\mu_i\epsi_{i11}\!-\!q_{i2}^2)\!-\!(\mu_i\epsi_{i13}\!+\!\xi q_{i2})\!(\mu_i\epsi_{i31}\!+\!\xi q_{i2})}} 
{\:\:\: q_{i1} \!\neq\! q_{i2}}
\\
\gamma_{i23}\!&=\! 
\twopartdef
{-\frac{\mu_i\epsi_{i32}}{\mu_i\epsi_{i33}-\xi^2}} 
{\qquad\qquad\quad\;\: q_{i1} \!=\! q_{i2}}
{-\frac{\mu_i\epsi_{i31}+\xi q_{i2}}{\mu_i\epsi_{i33}-\xi^2}\gamma_{i21}-\frac{\mu_i\epsi_{i32}}{\mu_i\epsi_{i33}-\xi^2}} 
{\qquad\qquad\quad\;\; q_{i1} \!\neq\! q_{i2}}
\\
%
%
%
%
\gamma_{i32}\!&= \!
\twopartdef
{0} 
{\qquad\quad\:\:\: q_{i3} \!=\! q_{i4}}
{\frac{\mu_i\epsi_{i21}(\mu_i\epsi_{i33}+\xi^2)-\mu_i\epsi_{i23}(\mu_i\epsi_{i31}+\xi q_{i3})}
{(\mu_i\epsi_{i33}-\xi^2)(\mu_i\epsi_{i22}-\xi^2-q_{i3}^2)-\mu_i^2\epsi_{i23}\epsi_{i32}}} 
{\qquad\quad\:\:\: q_{i3} \!\neq\! q_{i4}}
\\
\gamma_{i33}\!&= \!
\twopartdef
{\frac{\mu_i\epsi_{i31}+\xi q_{i3}}{\mu_i\epsi_{i33}-\xi^2}} 
{\qquad\qquad\qquad\:\, q_{i3}\! =\! q_{i4}}
{\frac{\mu_i\epsi_{i31}+\xi q_{i3}}{\mu_i\epsi_{i33}-\xi^2} +\frac{\mu_i\epsi_{i32}}{\mu_i\epsi_{i33}-\xi^2} \gamma_{i32}} 
{\qquad\qquad\qquad\:\, q_{i3} \!\neq\! q_{i4}}
\\
\gamma_{i41}\!&= \!
\twopartdef
{0} 
{\:\:\: q_{i3} \!=\! q_{i4}}
{\frac{\mu_i\epsi_{i32}(\mu_i\epsi_{i13}+\xi q_{i4})-\mu_i\epsi_{i12}(\mu_i\epsi_{i33}-\xi^2)}
{(\mu_i\epsi_{i33}\!-\!\xi^2)\!(\mu_i\epsi_{i11}\!-\!q_{i4}^2)\!-\!(\mu_i\epsi_{i13}\!+\!\xi q_{i4})\!(\mu_i\epsi_{i31}\!+\!\xi q_{i4})}} 
{\:\:\: q_{i3} \!\neq\! q_{i4}}
\\
\gamma_{i43}\!&= \!
\twopartdef
{-\frac{\mu_i\epsi_{i32}}{\mu_i\epsi_{i33}-\xi^2}} 
{\qquad\qquad\quad\;\: q_{i3} \!=\! q_{i4}}
{-\frac{\mu_i\epsi_{i31}+\xi q_{i4}}{\mu_i\epsi_{i33}-\xi^2}\gamma_{i41}-\frac{\mu_i\epsi_{i32}}{\mu_i\epsi_{i33}-\xi^2}} 
{\qquad\qquad\quad\;\: q_{i3} \!\neq\! q_{i4}.}
\end{split}
\end{align}

These solutions are finite and continuous for isotropic and anisotropic media, and therefore can be used to formulate a generalized, stable transfer matrix. 

For this purpose, the boundary conditions for electric and magnetic fields are applied in order to connect the fields of two adjacent layers $i-1$ and $i$.  Formulated for all four modes simultaneously, the boundary conditions in matrix form become:

\begin{align}
\AMa{i-1} \Evec_{i-1}=\AMa{i} \Evec_i,
\label{eq:A_boundaryShort}
\end{align}
where $\AMa{i}$ is calculated in terms of $\gamma_{ijk}$ by \cite{Xu2000}
\begin{align}
\AMa{i}\!\!=\!\!\fourfourMatrix
{\ga{i11} & \ga{i21} & \ga{i31} & \ga{i41}}
{\ga{i12} & \ga{i22} & \ga{i32} & \ga{i42}}
{\!\!\!\frac{q_{i1} \ga{i11}\!-\!\xi \ga{i13}}{\mu_i}
\!\!& \!\! \frac{q_{i2} \ga{i21}\!-\!\xi \ga{i23}}{\mu_i}
\!\!& \!\! \frac{q_{i3} \ga{i31}\!-\!\xi \ga{i33}}{\mu_i}
\!\!& \!\! \frac{q_{i4} \ga{i41}\!-\!\xi \ga{i43}}{\mu_i}\!\!\!}
{\frac{1}{\mu_i}q_{i1}\ga{i12}
& \frac{1}{\mu_i}q_{i2}\ga{i22}
& \frac{1}{\mu_i}q_{i3}\ga{i32}
& \frac{1}{\mu_i}q_{i4}\ga{i42}}\!\!,\!
\label{eq:A_Amatrix}
\end{align}
and $\Evec_{i}$ is the dimensionless 4-component electric field vector of the solution, i.e., representing the field for the full multilayer system. Conveniently,  $\Evec_{i}$ is defined as \cite{Xu2000}:
\begin{align}
\label{eq:E_xu_sorting}
\Evec\equiv\colvec{E_{\text{trans}}^p \\ E_{\text{trans}}^s \\ E_{\text{refl}}^p \\E_{\text{refl}}^s}
\end{align}
where $E_{\text{trans}}^{p(s)}$ and $E_{\text{refl}}^{p(s)}$ are the in-plane field amplitudes of the transmitted and reflected mode, p-polarized and s-polarized, respectively. 

By multiplying $\AMa{i}^{-1}$ on both sides of eq.~\ref{eq:A_boundaryShort}, the following relation is obtained, where the interface matrix \LMa{i}~is defined implicitly:
\begin{align}
\Evec_{i-1}=\AMa{i-1}^{-1}\AMa{i}\Evec_{i}\equiv \LMa{i}\Evec_{i}.
\end{align}

By inspection of eq.~\ref{eq:A_Amatrix}~it becomes clear that the columns of matrix \AMa{i}~ resemble the four eigenmodes $\Psi_{ij}$ of the respective layer, however, with the order of the elements being $E_x$, $E_y$, $H_x$, and $H_y$. Therefore, the matrix operation $\AMa{i}\Evec_{i}$ essentially represents the projection of the solution field vector onto the eigenmodes of the layer, while $\AMa{i-1}^{-1}\AMa{i}$ projects the eigenmodes in layer $i$ onto the ones of layer $i-1$. The different order of the column vector elements as compared to eq.~\ref{eq:A_reorderedFieldVec} must be accounted for, which will be done conveniently after formulating the transfer matrix of the complete multilayer system.

In order to do so, first the propagation matrix \PMa{i}, as it was called by Yeh \cite{Yeh1979}, has to be defined. For the sign convention chosen here, it is given by
\begin{align}
\PMa{i}\!=\!\fourfourMatrix
{e^{-i\frac{\omega}{c}q_{i1}d_i} & 0 & 0 & 0}
{0 & e^{-i\frac{\omega}{c}q_{i2}d_i} & 0 & 0}
{0 & 0 & e^{-i\frac{\omega}{c}q_{i3}d_i} & 0}
{0 & 0 & 0 & e^{-i\frac{\omega}{c}q_{i4}d_i}}\!.
\end{align}

With this, the transfer matrix \TMa{i}~of a single layer $i$, which is composed of parts from the enclosing interface matrices and the propagation matrix, is defined as:
\begin{align}
\TMa{i}=\AMa{i} \PMa{i} \AMa{i}^{-1},
\end{align}
and the transfer matrix \TMa{\text{tot}}~of all $N$ layers is then obtained by evaluating
\begin{align}
\TMa{\text{tot}}=\prod_{i=1}^{N} \TMa{i}.
\end{align}
However, \TMa{\text{tot}}~on its own is not sufficient to calculate the reflectivity or transmittance of the multilayer system, since for the first and the last interface, i.e. the interfaces with the incident medium $i=0$ and the substrate $i=N+1$, only half of the interface matrices \LMa{1}~and \LMa{N+1} are included. The full transfer matrix \FullTMa{N}~is obtained as follows, where different notations are shown in order to clarify the equivalence of the various approaches found in the literature \cite{Berreman1972,Lin-Chung1984,Xu2000,Yeh1979}:
\begin{align}
\!\!\!\begin{split}
\FullTMa{N} &=\! \AMa{0}^{-1} \TMa{\text{tot}} \AMa{N+1} \\
&=\! \AMa{0}^{-1} \TMa{1} \TMa{2} ... \TMa{N} \AMa{N+1} \\
&=\! \AMa{0}^{-1} \AMa{1} \PMa{1} \AMa{1}^{-1} \AMa{2} \PMa{2} \AMa{2}^{-1} ... \AMa{N} \PMa{N} \AMa{N}^{-1} \AMa{N+1} \\
&=\! \LMa{1} \PMa{1} \LMa{2} \PMa{2} ... \LMa{N} \PMa{N} \LMa{N+1}.
\end{split}
\end{align}

The first line is perfectly suited for the implementation in a computer program, and furthermore allows to directly see how the transfer matrix of a single interface ($N=0$) is calculated, i.e. $\FullTMa{0}=\AMa{0}^{-1}\AMa{1}$. The last line, on the other hand, illustrates how systematic the matrix approach solves the propagation of electromagnetic waves in a multilayer medium, simply stringing together interface matrices \LMa{i}~and propagation matrices \PMa{i}, for each interface and layer, respectively.

\begin{figure*}
\includegraphics[width = .9\textwidth]{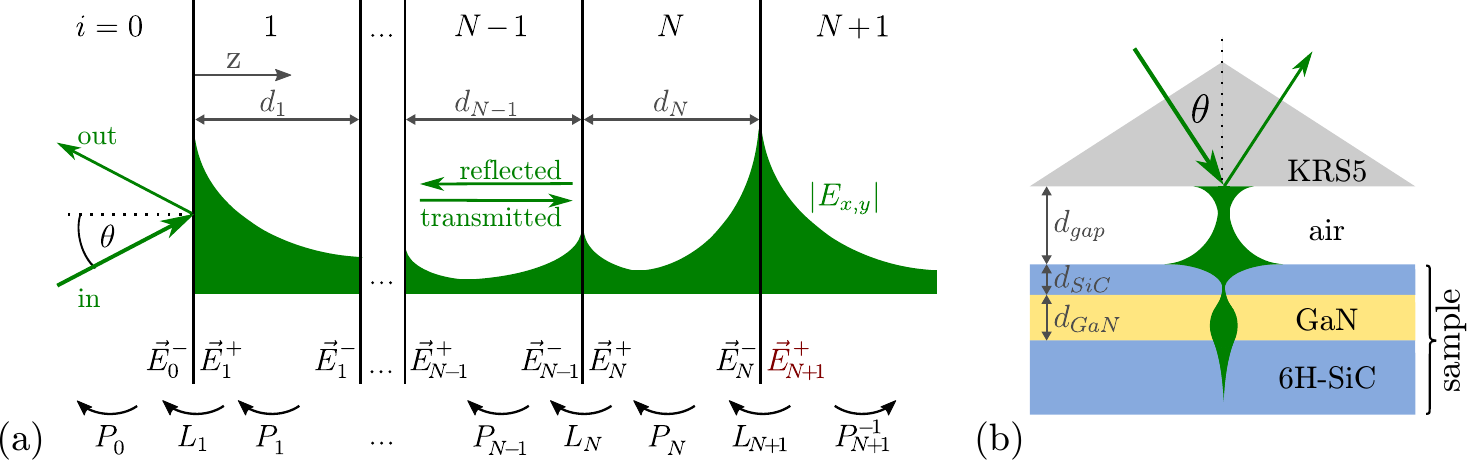}
\caption{(a) By means of the transfer-matrix formalism, the \Evec-field distribution can be calculated at any point in the multilayer system. Starting from $\Evec_{N+1}^+$, subsequent multiplication of the interface matrices \LMa{i} and propagation matrices \PMa{i} allows to propagate the wave back to the incident medium, and into the substrate. For the case of evanescent waves, the \Evec-field is exemplary sketched in green. (b) In the Otto geometry (not to scale), a highly refractive prism provides the necessary in-plane momentum for resonant coupling to  SPhPs, propagating along the interfaces of the multilayer system. Here, the SiC / GaN / SiC heterostructure is sketched, discussed in Sec.~3\ref{sec:sim.C}.}
\label{fig:multilayerStructure}
\end{figure*}

From the full transfer matrix \FullTMa{N}, reflection and transmission coefficients for both s- and p-polarized~incident, reflected, and transmitted waves can be calculated, as it has been shown by Yeh \cite{Yeh1979}, using:
\begin{align}
\Evec^-_0 = \FullTMa{N} \Evec^+_{N+1},
\end{align}
where $\Evec^-_{i-1}$ and $\Evec^+_i$ denote the fields on both sides of the interface between layer $i$ and $i-1$, respectively, see Fig.~\ref{fig:multilayerStructure} (a). However, before writing down the equations, the matrix \FullTMa{N}~has to be transformed such that the order of the field components corresponds to the chosen order of Yeh: 
\begin{align}
\Evec^*=\colvec{E_{\text{trans}}^p \\ E_{\text{refl}}^p \\ E_{\text{trans}}^s \\E_{\text{refl}}^s},
\end{align}
in contrast to the sorting shown in Eq.~\ref{eq:E_xu_sorting}, which the transfer matrix \FullTMa{N} is built from. Therefore, the transformation
\begin{align}
\YehTMa{N}=\Lambda_{1324}^{-1} \FullTMa{N} \Lambda_{1324},
\end{align}
where $\Lambda_{1324}$ is given by
\begin{align}
\Lambda_{1324}=\fourfourMatrix{1&0&0&0}{0&0&1&0}{0&1&0&0}{0&0&0&1},
\end{align}
yields the transfer matrix $\YehTMa{N}$ that is compatible with the Yeh formalism \cite{Yeh1979}.

\subsection{Reflectivity and Transmittance}
Employing the transformed transfer matrix of the complete multilayer system \YehTMa{N}, the reflection and transmission coefficients for s- and p-polarized~light, i.e. $r_{ss}$, $r_{pp}$, $t_{ss}$, and $t_{pp}$, and the mode coupling reflection and transmission coefficients $r_{sp}$, $r_{ps}$, $t_{sp}$, and $t_{ps}$ can be calculated, where the subscripts refer to incoming and outgoing polarization, respectively. The coefficients are given in terms of the matrix elements of \YehTMa{N} as follows \cite{Yeh1979}:
\begin{align}
r_{pp}&=\frac{\YTME{21}\YTME{33}-\YTME{23}\YTME{31}}{\YTME{11}\YTME{33}-\YTME{13}\YTME{31}} & 
t_{pp}&=\frac{\YTME{33}}{\YTME{11}\YTME{33}-\YTME{13}\YTME{31}} 
\label{eq:reflectionCoefficient_pp}
\\
r_{ss}&=\frac{\YTME{11}\YTME{43}-\YTME{41}\YTME{13}}{\YTME{11}\YTME{33}-\YTME{13}\YTME{31}} &
t_{ss}&=\frac{-\YTME{11}}{\YTME{11}\YTME{33}-\YTME{13}\YTME{31}} \\
r_{ps}&=\frac{\YTME{41}\YTME{33}-\YTME{43}\YTME{31}}{\YTME{11}\YTME{33}-\YTME{13}\YTME{31}} &
t_{ps}&=\frac{\YTME{31}}{\YTME{11}\YTME{33}-\YTME{13}\YTME{31}} \\
r_{sp}&=\frac{\YTME{11}\YTME{23}-\YTME{21}\YTME{13}}{\YTME{11}\YTME{33}-\YTME{13}\YTME{31}} &
t_{sp}&=\frac{\YTME{13}}{\YTME{11}\YTME{33}-\YTME{13}\YTME{31}},
\end{align}
and the reflectivity $R$ and intensity enhancement $T$ for a certain polarization are obtained by calculating the absolute square of the corresponding coefficient, i.e. $R_{kl} = |r_{kl}|^2$ and $T_{kl} = |t_{kl}|^2$ with $k,l=s,p$. Please note that the total transmission of the multilayer stack is distinct from $T$, where additionally energy conservation needs to be accounted for. In the case of an absorptive substrate, the transmission coefficients represent the magnitude of the in-plane component of the \Evec-field vector in the substrate at the interface with layer $N$. This particular fact will be clarified in the following, where we calculate the electric field distribution in the multilayer system employing the presented transfer-matrix formalism.

\subsection{Electric Field Distribution}

Using the interface and propagation matrices \LMa{i}~and \PMa{i}, respectively, an electric field vector can be projected to any point in the multilayer system. As a starting point, the transmission coefficients are utilized to formulate the in-plane components of the electric field $\Evec_{N+1}^+$ in the substrate at the interface with layer $N$, see Fig.~\ref{fig:multilayerStructure} (a):
\begin{align}
\Evec_{N+1}^+=\colvec{E_{\text{trans}}^p \\ E_{\text{trans}}^s \\ E_{\text{refl}}^p \\E_{\text{refl}}^s}=\colvec{t_{pp}+t_{sp} \\ t_{ss}+t_{ps} \\ 0 \\ 0},
\end{align}
where the reflected components are set to zero, since no light source is assumed to be on the substrate side of the multilayer system. Please note that we here conveniently go back to the field sorting according to eq.~\ref{eq:E_xu_sorting}, allowing to directly reuse any interface and propagation matrices that were calculated along the way to the full transfer matrix \FullTMa{N}. As it is shown in Fig.~\ref{fig:multilayerStructure} (a), the electric field vectors on both sides of an interface are connected by the interface matrix $\Evec^-_{i-1}=\LMa{i} \Evec_{i}^+$. Furthermore, the propagation through layer $i$ is given by the propagation matrix \PMa{i}, which can be evaluated $z$-dependently:
\begin{align}
\begin{split}
\Evec_{i}(z)&=\PMa{i}(z) \Evec^-_{i} \\
&=\!\fourfourMatrix
{\!e^{-i\frac{\omega}{c}q_{i1}z}\! & 0 & 0 & 0}
{0 & \!e^{-i\frac{\omega}{c}q_{i2}z}\! & 0 & 0}
{0 & 0 & \!e^{-i\frac{\omega}{c}q_{i3}z}\! & 0}
{0 & 0 & 0 & \!e^{-i\frac{\omega}{c}q_{i4}z}\!}\!\Evec^-_{i},
\end{split}
\end{align}
with $0<z<d_i$ being the relative $z$-position in layer $i$. As it is indicated by the black arrows in Fig.~\ref{fig:multilayerStructure}, starting from $\Evec_{N+1}^+$, interface matrices \LMa{i}~and propagation matrices \PMa{i}~are used to subsequently propagate the wave towards the incident medium. In the reverse direction, the inverse propagation matrix $\PMa{N+1}^{-1}$ allows to calculate the \Evec-fields in the substrate. 

By this means, the four in-plane components $E_{\text{trans}}^p$, $E_{\text{trans}}^s$, $E_{\text{refl}}^p$, and $E_{\text{refl}}^s$ are obtained as a function of $z$. Within each layer, there are only two possible shapes of the absolute amplitude of these components: they either describe a (damped) sinusoidal propagating wave, or an exponentially decaying evanescent wave. In order to get the complete field present in the multilayer system, the forward (transmitted) and backward (reflected) components for each polarization have to be added up. This summation of the components is performed as follows:
\begin{align}
E_x&=E_{\text{trans}}^p-E_{\text{refl}}^p  \\
E_y&=E_{\text{trans}}^s+E_{\text{refl}}^s  \\
E_z&=\left( -\frac{\xi \epsi_x}{q_{i1}\epsi_z} E_{\text{trans}}^p \right) - \left( -\frac{\xi \epsi_x}{q_{i3}\epsi_z} E_{\text{refl}}^p \right), 
\label{eq:Ez} 
\end{align}
where the negative signs for $E_x$ and $E_z$ account for the phase flip during reflection. The out-of-plane component $E_z$ has been calculated employing Maxwell's Eq.~$\nabla \times \Hvec=\epsi \frac{d\Evec}{dt}$ for $E_{\text{trans}}^p$ and $E_{\text{refl}}^p$ individually. The corresponding out-of-plane wave vector components $q_z$ are given by the layer-dependent solutions $j=1$ and $j=3$, i.e. $q_{\text{trans}}^p=q_{i1}$ and $q_{\text{refl}}^p=q_{i3}$, as defined in Eqs.~\ref{eq:sorting1} and \ref{eq:sorting2}. Note that Eq.~\ref{eq:Ez} is only valid for a dielectric tensor that is diagonal in the lab frame.

In the case of evanescent waves, the \Evec~field components typically have peaks at the interfaces and valleys inside the layers, which is sketched in Fig.~\ref{fig:multilayerStructure}~(a) in green. Such evanescent waves can occur for strong absorption in the medium or, alternatively, if the incident angle $\theta$ is larger than the critical angle of total internal reflection from the incidence medium to medium $i=1$, implying that the refractive indices fulfill $n_0>n_1$. This is the case for resonant excitation of SPhPs in the Otto geometry \cite{Falge1973,Neuner2009,Passler2017}, being analyzed in section \ref{sec:simulations}.

\subsection{Distinction to Previous Algorithms}
In the last decades, many authors have provided different $4 \times 4$ matrix approaches for the calculation of light propagation in anisotropic stratified media, but either they focus on special cases, thus lacking full generality, or the calculation, when implemented in a computer program, leads to numerical instabilities. For instance, the Berreman and Yeh formalisms \cite{Berreman1972,Yeh1979} as well as the generalized approach by Lin-Chung and Teitler \cite{Lin-Chung1984} assume solely fully anisotropic tensors, and hence lead to singularities if the material is isotropic or even if the three main dielectric components coincide with the laboratory coordinate system. A solution to these singularities is given by Xu \cite{Xu2000}, who, on the other hand, provides an unstable solution for solving the eigensystem of electric field modes of the multilayer system. Neither of these approaches delivers a robust strategy for uniquely assigning the different modes in each layer, as given by Li \textit{et al.} \cite{Li1988}. Here, by combining the different approaches, a numerically stable formalism is achieved, which is capable of handling isotropic or anisotropic, as well as absorbing or non-absorbing materials for each layer including the substrate. 

Specifically, we calculate the four eigenmodes of each layer as derived by Berreman \cite{Berreman1972}, which are correctly sorted into forward and backward, s- and p-polarized rays using the approach of Li \textit{et al.} \cite{Li1988}. The four eigenvalues $q_{ij}$ are used to obtain interface matrices free from discontinuities as shown by Xu \textit{et al.} \cite{Xu2000}. Following the generalized approach of Lin-Chung \textit{et al.}~\cite{Lin-Chung1984}, we then compose the complete transfer matrix correctly, and finally specify the transformation necessary for the calculation of reflection and transmission coefficients based on the work of Yeh \cite{Yeh1979}. In addition, we specifically illustrate how to calculate the electric field distribution throughout the full heterostructure. The formalism presented here is tailored for the robust calculation of light propagation in any isotropic or anisotropic stratified medium. Its generalized form is mandatory for the simulations involving highly dispersive and strongly anisotropic dielectric permittivities, as it is the case for SPhPs in polar dielectric heterostructures, being discussed in the following section.

\section{Simulations}
\label{sec:simulations}

In general, the transfer-matrix formalism presented in the previous section can be employed for any wavelength and any number of layers, each described by an arbitrary dielectric tensor. In the following, we focus on SPhPs in polar crystals in the mid-infrared spectral region. These surface modes exist due to negative dielectric permittivity in the Reststrahlen band between the transverse optical (TO) and longitudinal optical (LO) frequencies \cite{Falge1973}. All cases discussed in the following are also implemented in the Matlab example code \cite{Passler2017a}.

Specifically, we employ our formalism for the simulation of SPhP resonances in four different model systems:  bare 6H-silicon carbide (SiC) in Sec.~\ref{sec:sim.A}, a thin layer of gallium nitride (GaN) on top of 6H-SiC in Sec.~\ref{sec:sim.B}, an additional thin SiC layer on top of GaN/SiC in Sec.~\ref{sec:sim.C}, as sketched in Fig.~\ref{fig:multilayerStructure}~(b), and $\alpha$-quartz in Sec.~\ref{sec:sim.D}. We chose these model heterostructures built from anisotropic, polar dielectrics in order to briefly demonstrate the variety of SPhP modes than can exist in such stratified systems, as a starting point for future detailed studies of the SPhPs in these and similar structures.

All materials considered here are uniaxial, and for Sections~\ref{sec:sim.A},~\ref{sec:sim.B}, and~\ref{sec:sim.C}, the crystal axes are chosen to be normal to the sample surface, i.e. $\epsiTens=\left( (\epsiPerp,0,0),(0,\epsiPerp,0),(0,0,\epsiPar) \right)$ with the ordinary and extraordinary dielectric functions, $\epsiPerp$ and $\epsiPar$, respectively, taken from literature \cite{Paarmann2016,Kasic2000}. In Sec.~\ref{sec:sim.D}, we simulate the dispersion of SPhPs for different crystal orientations of $\alpha$-quartz \cite{Gervais1975}, being a natural hyperbolic material \cite{RodriguesdaSilva2010} in the frequency range of \wavenumber{350-600}. For all schemes, the SPhPs are excited in the Otto geometry \cite{Falge1973,Neuner2009} using KRS5 as a highly refractive coupling medium ($n\approx 2.4$). Using p-polarized light at incidence angles above the critical angle of total internal reflection ($\theta_{\text{crit}} \approx$ 25$^\circ$), the large in-plane momentum of the evanescent wave in the air gap allows to couple to the SPhP modes in the multilayer structure \cite{Passler2017}.

\begin{figure*}
\includegraphics[width = \textwidth]{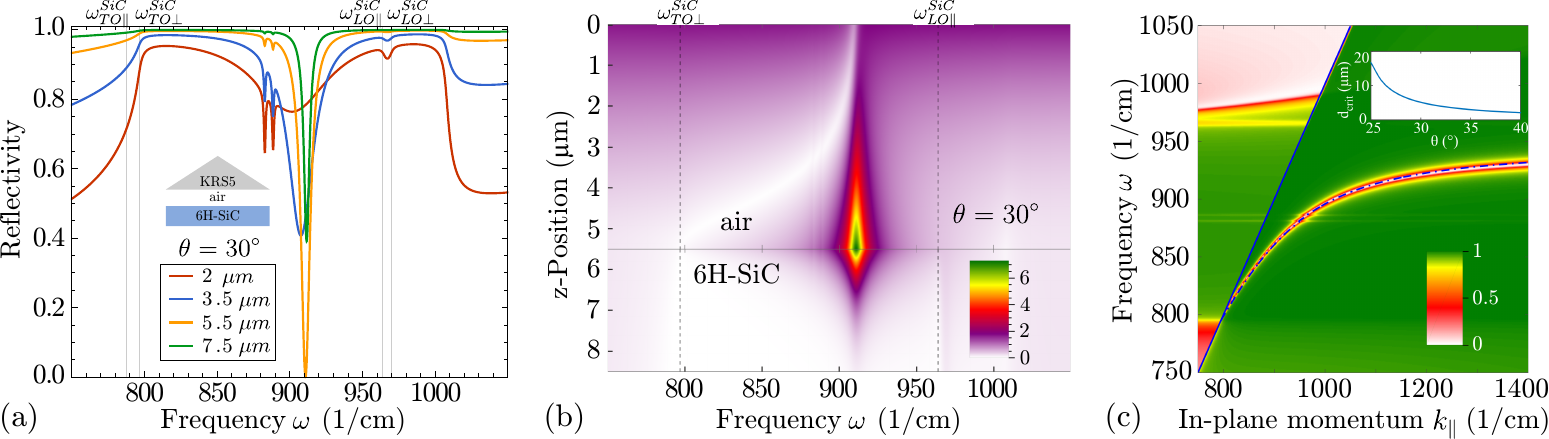}
\caption{Simulations of a single SPhP resonance in 6H-SiC excited in the Otto geometry. (a) Reflectivity spectra in the Reststrahlen region at \dg{30} incident angle for four different gap widths, illustrating the critical coupling behavior of the SPhP resonance at $\sim\wavenumber{910}$. (b) $E_x$-field enhancement at the air-SiC interface as a function of excitation frequency and $z$-position. The simulation reveals the spatial localization of the field enhancement of the SPhP and allows to predict the non-linear yield, see text. (c) Simulated reflectivity map showing the SPhP dispersion in the three-layer Otto configuration together with the theoretical two-layer dispersion (dashed-dotted blue line), both being in excellent agreement.}
\label{fig:SiC}
\end{figure*}  

\subsection{Surface Phonon Polariton in 6H-SiC}
\label{sec:sim.A}

In a bare SiC sample, a single SPhP mode propagating along the air-SiC interface is known to be accessible at frequencies inside the Reststrahlen region between $\omega_{TO,\perp}=\wavenumber{797}$ and $\omega_{LO,\perp} =\wavenumber{968}$ \cite{Neuner2009,Passler2017}. In the Otto geometry \cite{Otto1968}, the air gap width $d_{\text{gap}}$ is a critical coupling parameter in terms of excitation efficiency, while the incidence angle $\theta$ allows to select the excitation frequency along the SPhP dispersion \cite{Passler2017}. Employing the transfer-matrix formalism, the reflectivity of such a system can be evaluated by means of Eq.~\ref{eq:reflectionCoefficient_pp} as a function of incidence frequency $\omega$, incidence angle $\theta$, and gap width $d_{\text{gap}}$. 

Four exemplary reflectivity spectra are shown in Fig.~\ref{fig:SiC} (a) at an incident angle of $\theta=\dg{30}$. The simulations feature the Reststrahlen band with high reflectivity between TO and LO frequencies, as well as the SPhP resonance at $\sim\wavenumber{910}$. The amplitude and width of the SPhP dip strongly depend on the air gap width,  demonstrating the critical coupling behavior \cite{Passler2017}. Furthermore, an anisotropy dip at the axial LO frequency \cite{Paarmann2016}, and the two weak modes arising due to zone-folding along the $c$-axis in 6H-SiC \cite{Bluet1999} are reproduced.

In Fig.~\ref{fig:SiC} (b), the absolute $\Evec_x$-field amplitude is shown as a function of frequency and $z$-position at $\theta=\dg{30}$ and the corresponding critical gap of $d_{\text{gap}}=\mumetr{5.5}$. The 6H-SiC substrate extends into the $z>\mumetr{5.5}$ half-space, while the prism lies at $z<0$, which is not shown in this plot. At the SPhP resonance frequency, a strong $\Evec_x$-field enhancement localized at the air-SiC interface is observed. As illustrated in a previous work \cite{Passler2017}, the field enhancement at the interface can be extracted to calculate the enhancement of second harmonic yield at the SPhP resonance in such a configuration. In order to determine the non-linear sources in a more complex structure, the \Evec-field distribution inside the multilayer system is a key tool. 

In Fig.~\ref{fig:SiC} (c), an axis-transformed reflectivity map is shown, exhibiting the SPhP dispersion at wave vectors exceeding the light line (solid blue line). The map is built by evaluating the reflectivity as a function of light frequency $\omega$ and incidence angle $\theta$, at the critical coupling gap width $d_{\text{crit}}$ for each $\theta$ \cite{Passler2017}, which is  plotted in the inset. For $\theta$ smaller than the angle of total internal reflection in the prism, the reflectivity is evaluated for $d_{\text{gap}}=0$. The numerical simulations perfectly agree with the theoretically calculated single-interface dispersion curve at the air-SiC interface (dashed-dotted blue line) \cite{Raether1988}. Following this procedure, our matrix formalism can be employed to predict polariton modes in more complicated structures and analyze their dispersion, as illustrated in the following examples. 

\begin{figure*}
\includegraphics[width = \textwidth]{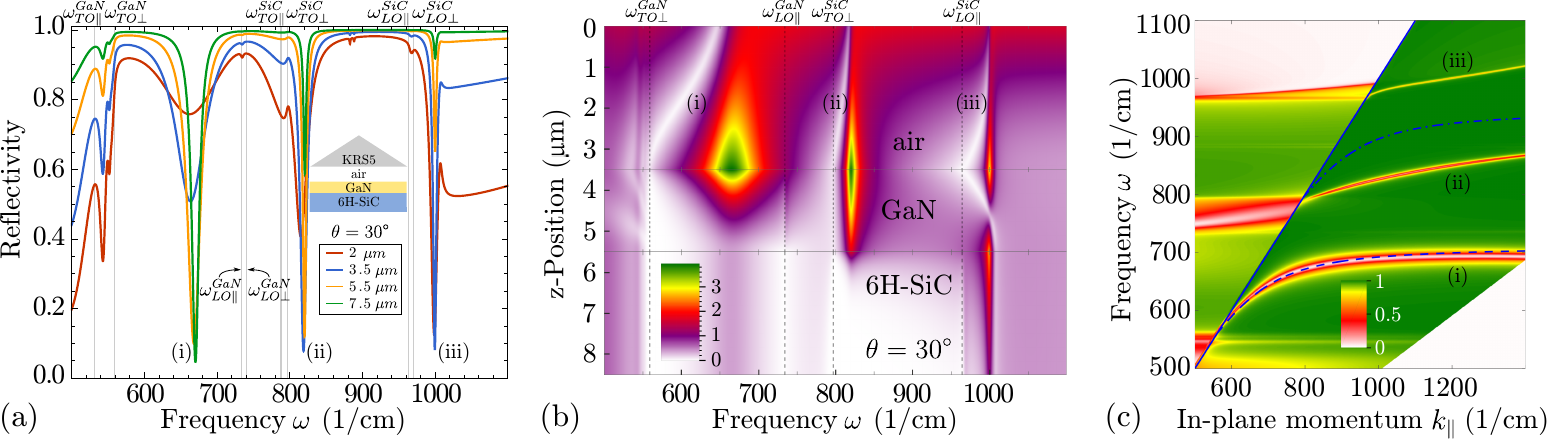}
\caption{Simulations of resonances in \mumetr{2}~GaN on 6H-SiC excited in the Otto geometry. (a) Reflectivity spectra in the Reststrahlen regions of both media at $\theta=\dg{30}$ for four different gap widths, illustrating the appearance of three modes. (b) Absolute $\Evec_x$-field amplitude in the air-GaN-SiC system as a function of excitation frequency and $z$-position. The simulation reveals the spatial localization of each mode: (i) the GaN SPhP mode at $\sim\wavenumber{680}$ is localized at the air-GaN interface, (ii) the index-shifted SiC-SPhP-like mode at $\sim\wavenumber{820}$ extends into the SiC substrate, and (iii) the waveguide-like mode at $\sim\wavenumber{1000}$ peaks at both interfaces, featuring a phase flip inside the GaN layer. (c) Simulated dispersions in the four-layer Otto configuration together with the theoretical two-layer SPhP dispersions of SiC (dashed-dotted blue line) and GaN (dashed blue line). The GaN SPhP dispersion is in excellent agreement with the simulations, indicating no significant modification by the SiC substrate. The SiC SPhP dispersion, on the other hand, is strongly index-shifted to lower frequencies by the presence of the GaN layer.}
\label{fig:GaNSiC}
\end{figure*}

\subsection{GaN / 6H-SiC}
\label{sec:sim.B}

In the second system we consider, a  \mumetr{2} thick film of hexagonal gallium nitride (GaN) is added on top of 6H-SiC, as shown in Fig.~\ref{fig:GaNSiC}. Note that the Reststrahlen band of GaN lies at lower frequencies between $\omega_{TO,\perp}=\wavenumber{560}$ and $\omega_{LO,\perp} =\wavenumber{742}$ \cite{Kasic2000} and does not overlap with the SiC Reststrahlen region. In the reflectivity spectra in Fig.~\ref{fig:GaNSiC} (a), three strong resonances can be observed at (i) $\sim\wavenumber{680}$, (ii) $\sim\wavenumber{810}$, and (iii) $\sim\wavenumber{1000}$. All resonances exhibit the SPhP-characteristic critical coupling behavior \cite{Passler2017,Neuner2009}, although each with a distinct critical gap width. Mode (i) appears clearly in the GaN Reststrahlen band. Mode (ii) is as a red-shifted SiC SPhP, the magnitude of red-shifting increasing with increasing GaN layer thickness (not shown), which can be interpreted as index-shifting of the SiC SPhP by the GaN layer \cite{Neuner2010}. Mode (iii) surprisingly emerges at the upper edge of the SiC high-reflectivity band at $\sim\wavenumber{1000}$, i.e., outside the spectral region supporting SPhPs, and likely corresponds to a waveguiding mode at small positive $\epsi_{\text{SiC}}$ and large $\epsi_{\text{GaN}}$ at this frequency.

The origin of the modes can be further analyzed by means of the $\Evec_x$-field distribution, which is calculated for $d_{\text{gap}}=\mumetr{3.5}$ and shown in Fig.~\ref{fig:GaNSiC} (b). Resonance (i) is localized at the air-GaN interface and is spectrally broad, since for  $d_{\text{gap}}=\mumetr{3.5}$, the excitation is over-coupled with large radiative losses back into the prism \cite{Passler2017}. Due to the negative permittivity of the GaN Reststrahlen band, the underlying SiC substrate has no considerable influence on this air-GaN SPhP. For mode (ii), on the other hand, the GaN layer is transparent and the SiC substrate exhibits negative permittivity. This polariton mode has its peak intensity at the air-GaN interface, but extends down into SiC. Mode (iii), on the other hand, has field enhancement peaks at both the air-GaN and the GaN-SiC interface and exhibits a phase flip inside the GaN layer, suggesting the first-order waveguiding nature of the mode. 

In the dispersion plot in Fig.~\ref{fig:GaNSiC} (c), the three polariton branches can be observed on the right-hand side of the light line. As a blue dashed line, the theoretical air-GaN SPhP dispersion is shown \cite{Raether1988}, coinciding with the lowest-frequency resonance identified as a GaN SPhP. The small deviation of the dispersion from the simulations at large wavenumbers arises due to calculating the map using the critical gap width of SiC (see inset in Fig.~\ref{fig:SiC} (c)), which is somewhat larger for GaN. The theoretical air-SiC SPhP dispersion is shown again (blue dot-dashed line) in order to demonstrate the large index-red-shift of the polariton mode (ii) due to the GaN layer. In contrast to the typical converging shape of the SPhP dispersion, mode (iii) exhibits an almost linear behavior, following the position of the $\theta$-dependent upper edge of the high-reflectivity band \cite{Passler2017}. In particular the unusual dispersion of the modes (ii) and (iii) is very interesting, and certainly deserves further systematic study. 

\begin{figure*}
\includegraphics[width = \textwidth]{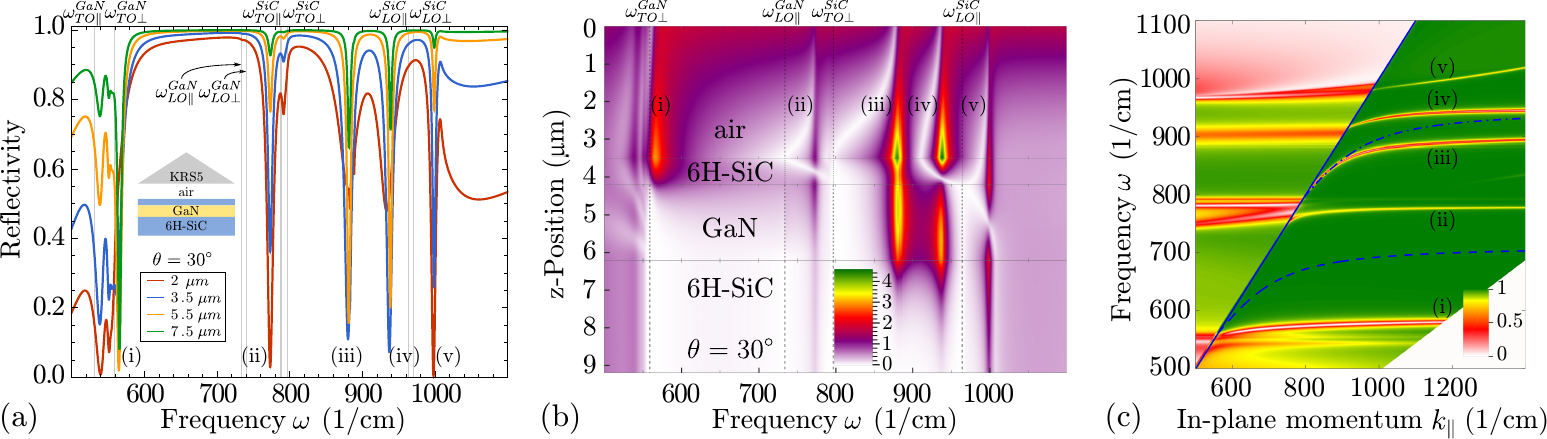}
\caption{Simulations of resonances in \nmetr{700}~6H-SiC  / \mumetr{2}~GaN on 6H-SiC excited in the Otto geometry. (a) Reflectivity spectra in the Reststrahlen region at \dg{30} incident angle for four different gap widths, showing the appearance of five different modes, labeled (i)-(v). (b) Absolute $\Evec_x$-field amplitude in the air-SiC-GaN-SiC system as a function of excitation frequency and $z$-position. The simulation reveals the spatial localization of each mode. The SiC SPhP appears to split into two branches (iii) and (iv) at $\sim\wavenumber{900}$, exhibiting symmetric and anti-symmetric field distributions in the SiC top film, coupled to a waveguiding mode in the GaN film. (c) Simulated dispersions in the five-layer Otto configuration together with the theoretical two-layer SPhP dispersions of SiC (dashed-dotted blue line) and GaN (dashed blue line). In contrast to the previous examples, all mode branches differ strongly from the respective single-interface SPhP dispersions.}
\label{fig:SiCGaNSiC}
\end{figure*} 

\subsection{6H-SiC / GaN / 6H-SiC}
\label{sec:sim.C}

On top of the previous system, we here consider an additional \nmetr{700} thick 6H-SiC layer, as sketched in Fig.~\ref{fig:multilayerStructure} (b). In Fig.~\ref{fig:SiCGaNSiC} (a), the reflectivity spectra of such a system are shown at $\theta=\dg{30}$, again for the same four different values of $d_{\text{gap}}$. As can be inferred from the critical coupling behavior accompanied by almost-zero reflectivity, the multilayer structure now supports at least five different modes, each with an individual critical gap width. 

Analogous to mode (iii) in Sec.~\ref{sec:sim.B}, a mode (v) appears at $\sim\wavenumber{1000}$, but here additionally a similar mode (ii) is observed at the upper GaN Reststrahlen edge at $\sim\wavenumber{750}$. By inspection of the \Evec-field distribution in Fig.~\ref{fig:SiCGaNSiC} (b), this similar nature of modes (ii) and (v) can be identified, both having field maxima at the interfaces of the GaN and SiC film, respectively, while exhibiting a phase flip inside the respective layer. Interestingly, mode (v), which spectrally appears at the upper SiC Reststrahlen edge, is spatially localized at the interfaces of the GaN layer, while mode (ii) occuring at $\sim\wavenumber{750}$, i.e. at the upper GaN Reststrahlen edge, is spatially localized at the SiC film interfaces. In consequence, the appearance of both modes can be attributed to a waveguide configuration with refractive indices $n_1 / n_2 / n_3$, where $n_2>n_1,n_3$. Additionally, the GaN SPhP (i) is now strongly modified by the SiC top layer, exhibiting a pronounced red-shift as compared to bare GaN.

The strongest field enhancement at the chosen gap of $d_{\text{gap}}=\mumetr{3.5}$, however, arises from the two resonances around \wavenumber{900} in Fig.~\ref{fig:SiCGaNSiC}~(b). These modes (iii) and (iv) are localized at the air-SiC interface, but extend further into the system with a second maximum inside the GaN layer. The field distribution in the SiC top layer suggests assignment to symmetric (iii) and asymmetric (iv) thin-film modes \cite{Berini2009}, with the asymmetric mode (iv) exhibiting a phase flip. However, the second enhancement inside the GaN layer points towards a GaN waveguided mode coupled to these SiC thin film modes. Further systematic studies are required to clarify the exact nature of these modes.

As for the previous systems, in Fig.~\ref{fig:SiCGaNSiC}~(c), the dispersion relations are shown. Clearly, the GaN SPhP, mode (i), is strongly index-shifted by the SiC overlayer owing to the large refractive index of SiC below its Reststrahlen band \cite{Paarmann2016}, as compared to the calculated SPhP on bare GaN (blue, dashed line). In fact, it is shifted so strongly that it is almost pushed out of the Reststrahlen band of GaN. Slightly thicker SiC films ($>\nmetr{850}$) result in complete quenching of the resonance (not shown). Despite their apparent similarity, modes (ii) and (v) show qualitatively different dispersions, converging for mode (ii) and diverging for mode (v). Possibly, this can be due to the adjacent lower SiC Reststrahlen edge for mode (ii), functioning as a hard upper limit for the dispersion. Finally, the dispersions of modes (iii) and (iv) enclose the single-interface SiC SPhP (blue, dash-dotted line), which is approached by the symmetric mode close to the light line, and by the asymmetric mode for large in-plane momenta. Note that due to the particular multilayer structure, the dispersions are clearly very different from the thin-film modes supported by a free-standing film \cite{Berini2009}.

The dispersion map shown in Fig.~\ref{fig:SiCGaNSiC}~(c) illustrates the rich polariton structure that emerges even for relatively simple multilayer systems. A full analysis of these modes clearly goes beyond the scope of this work, but we here show that our algorithm provides a complete toolset for further studies.

\subsection{Surface Phonon Polaritons in Hyperbolic $\boldsymbol{\alpha}$-Quartz}
\label{sec:sim.D}

\begin{figure*}
\includegraphics[width = \textwidth]{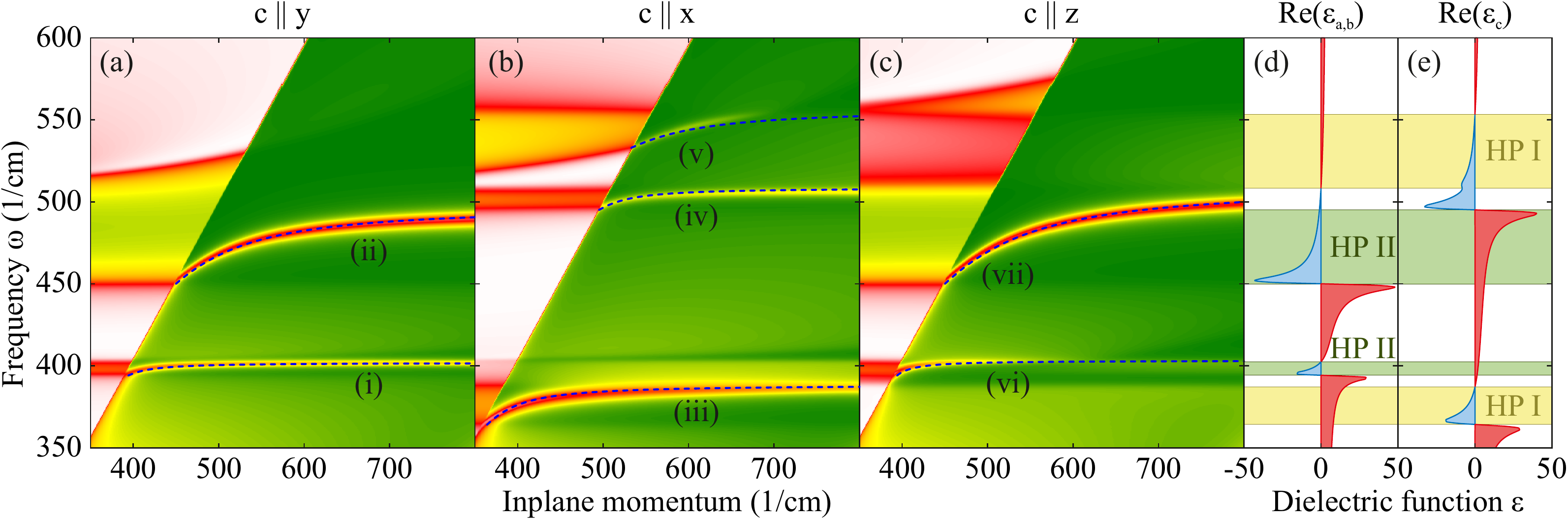}
\caption{Simulations of SPhP dispersions in hyperbolic $\alpha$-quartz excited in the Otto geometry. (a), (b), and (c) show critical-gap reflectivity maps for the crystal orientations with $c \parallel y$, $c \parallel x$, and $c \parallel z$, respectively. The critical gap has been estimated by taking $2d_{\text{crit}}$ from SiC, accounting for the increased penetration depth of evanescent waves at half the wavelength \cite{Passler2017}. Nonetheless, the modes exhibit different critical coupling conditions due to different damping constants. As blue dashed lines, the two-layer SPhP dispersions\cite{Falge1973} are shown, being in excellent agreement with the simulations. In (d) and (e), the real part of the ordinary ($\epsi_{a,b}$) and extraordinary ($\epsi_c$) dielectric components are shown, plotted in red and blue for the positive and negative regions, respectively, for better visibility. In yellow and green we highlight the regions where $\alpha$-quartz is hyperbolic of type I and II, respectively, allowing for direct identification of the modes observed in (a-c).}
\label{fig:quartz}
\end{figure*} 

The previous examples dealt with materials exhibiting a relatively small anisotropy, where the axial-planar phonon frequency difference is smaller than the TO-LO splitting. There is a number of materials having a large anisotropy instead. For example, h-BN exhibits very large optical phonon anisotropy such that the planar (ordinary) and axial (extraordinary) Reststrahlen bands do not overlap anymore \cite{Dai2014,Yoxall2015,Caldwell2014, Jacob2014, Li2015}. Similarly, several oxides like $\alpha$-quartz or sapphire exhibit different numbers of axial and planar infrared-active phonon modes \cite{Gervais1975,Gervais1974}. Both scenarios can lead to infrared spectral regions where the material exhibits a hyperbolic dispersion of its dielectric response, meaning that the real part of its dielectric tensor has both positive and negative principle components \cite{Jacob2014,Caldwell2014}. In the spectral regions where only one element is negative, the hyperbolic material is of type I, while for two negative components, it is of type II \cite{Caldwell2014}.  

As an alternative natural hyperbolic material \cite{RodriguesdaSilva2010}, we here simulate the polariton dispersions in $\alpha$-quartz in the region of \wavenumber{350-600}. A systematic description of all crystal orientation-dependent SPhP modes was presented by Falge and Otto \cite{Falge1973}, whose two-layer dispersion formulas are employed for comparison with our transfer-matrix simulations. In Fig.~\ref{fig:quartz} (a), (b), (c) we show reflectivity maps in the Otto geometry for three crystal orientations with the $c$-axis parallel to $x$, $y$ (both in-plane), and $z$ (out-of-plane), respectively. The blue dashed lines, representing the two-layer dispersions from Ref.~\cite{Falge1973}, are in excellent agreement with the simulated dispersion curves obtained from the transfer-matrix formalism. In Fig.~\ref{fig:quartz} (d) and (e), the real part of the dielectric function along the ordinary axes ($\epsi_{a,b}$) and along the extraordinary $c$-axis ($\epsi_c$), respectively, are shown. In yellow and green, the regions are marked where $\alpha$-quartz is type I (HPI) and type II (HPII) hyperbolic, respectively. 

For $c \parallel y$, the two appearing SPhPs originate from the ordinary dielectric components $\epsi_{a,b}$ only because of their intrinsic p-polarized nature, and hence are called "ordinary" SPhPs \cite{Falge1973}. In the case of an $a$-cut crystal ($c \parallel x$), on the other hand, one normal SPhP ($\epsi_x,\epsi_z<0$) at $\approx\wavenumber{500}$ and two modes in the HPI regions exist. This situation is reversed in a $c$-cut crystal ($c \parallel z$), where two modes appear in the HPII regions. The upper mode, however, reaches up into the normal SPhP region around \wavenumber{500}, thereby changing its character along the dispersion. Note that we here only observe modes for the in-plane dielectric component $\epsi_x$ being negative, while the out-of-plane component can be negative (normal) or positiv (hyperbolic). 

In contrast to hBN, $\alpha$-quartz can be fabricated with any crystal orientation, therefore enabling the possibility of custom-designed polariton resonances. However, its intrinsic hyperbolic nature and the character of the appearing SPhP modes have not yet been investigated in detail. As we have shown, the presented transfer-matrix formalism proves to be an excellent tool for the analysis of the linear response of highly anisotropic materials such as $\alpha$-quartz, allowing for quantitative predictions for polariton modes.

\subsection{Discussion}
We have presented examples of anisotropic multilayer structures made from polar dielectrics reaching into sub-wavelength dimensions, as well as phonon polariton resonances in hyperbolic materials. Combining these two can lead to effective material properties with yet unexplored functionality and tunability. As the layer thicknesses $d$ are further reduced, extreme optical confinement $\lambda/d > 1000$ may be achieved which is typically inaccessible in plasmonics due to the much shorter wavelengths employed \cite{Campione2015}. In the ultimate regime of atomic-scale heterostructures \cite{Caldwell2016,Woessner2015}, one can additionally expect the optical properties to deviate from calculations using bulk parameters of the constituent materials due to microscopic modification of the material properties \cite{Caldwell2016}. As we have shown, the formalism presented here is perfectly suited to simulate the optical response of such systems and could therefore be very useful in future studies of ultrathin-film dielectric heterostructures. 

Additionally, our formalism can be straight-forwardly extended into the nonlinear regime with arbitrary distribution of nonlinear optical sources. Notably, nonlinear-optical effects such as second harmonic generation (SHG) or sum-frequency generation are known to be extremely sensitive to the layer arrangements and thicknesses in thin-film heterostructures \cite{Razdolski2016b,Palomba2008,OBrien2013}. For SPhPs, only a few proof-of-concept nonlinear experiments have been performed \cite{Razdolski2016,Passler2017}, demonstrating that the sub-diffractional light localization can lead to drastic enhancement of the SHG yield. It is then obvious that thin-film heterostructures of polar dielectrics would allow tuning of the mid-infrared nonlinear properties with yet unexplored level of control and enhancement. The algorithm presented here provides the complete electric field distributions in such structures, enabling predictive studies of the nonlinear-optical response.

\section{Conclusion}

We have presented a generalized $4\times 4$ matrix formalism allowing to calculate the linear optical response of arbitrarily anistropic or isotropic, absorbing or non-absorbing multilayer structures. The algorithm is comprehensible, robust, free of discontinuous solutions, and can easily be implemented in a computer program \cite{Passler2017a}. The robustness of the algorithm is achieved by combining previous formalisms such that discontinuities and poles are completely avoided in the different steps of the algorithm. We give the equations for reflection and transmission coefficients, as well as for the full electric field distribution throughout the heterostructure. As a test ground, we applied the algorithm to simulations of SPhPs excited in the Otto geometry for selected anisotropic multilayer samples, where we observed critical coupling behavior, index shifting, mode splitting, wave-guiding, as well as phonon polaritons in hyperbolic materials. Our algorithm holds high promise for predicting SPhP modes in ultrathin films, hyperbolic structures, and atomic-scale heterostructures, as well as their nonlinear-optical response.

\section{Acknowledgments} 
We thank I. Razdolski (FHI Berlin) for careful reading of the manuscript, and M. Wolf (FHI Berlin) for supporting this work.


\begin{thebibliography}{10}
\newcommand{\enquote}[1]{``#1''}

\bibitem{Maier2007}
S.~A. Maier, \emph{{Plasmonics: Fundamentals and applications}} (Springer US,
  2007).

\bibitem{Hiller2002}
J.~Hiller, J.~D. Mendelsohn, and M.~F. Rubner, \enquote{{Reversibly erasable
  nanoporous anti-reflection coatings from polyelectrolyte multilayers},}
  Nature Materials \textbf{1}, 59--63 (2002).

\bibitem{Xi2007}
J.-Q. Xi, M.~F. Schubert, J.~Kim, E.~F. Schubert, M.~Chen, S.-Y. Lin, W.~Lie,
  and J.~A. Smart, \enquote{{Optical thin-filmmaterials with low refractive
  index for broadband elimination of Fresnel reflection},} Nature Photonics
  \textbf{1}, 176 (2007).

\bibitem{Nomura2003}
K.~Nomura, \enquote{{Thin-Film Transistor Fabricated in Single-Crystalline
  Transparent Oxide Semiconductor},} Science \textbf{300}, 1269--1272 (2003).

\bibitem{Street2009}
R.~A. Street, \enquote{{Thin-Film Transistors},} Advanced Materials
  \textbf{21}, 2007--2022 (2009).

\bibitem{Chopra2004}
K.~L. Chopra, P.~D. Paulson, and V.~Dutta, \enquote{{Thin-film solar cells: an
  overview},} Progress in Photovoltaics: Research and Applications \textbf{12},
  69--92 (2004).

\bibitem{Shin2011}
B.~Shin, O.~Gunawan, Y.~Zhu, N.~A. Bojarczuk, S.~J. Chey, and S.~Guha,
  \enquote{{Thin film solar cell with 8.4{\%} power conversion efficiency using
  an earth-abundant Cu 2 ZnSnS 4 absorber},} Progress in Photovoltaics:
  Research and Applications \textbf{21}, 72--76 (2013).

\bibitem{Patel2003}
N.~Patel, P.~Patel, and V.~Vaishnav, \enquote{{Indium tin oxide (ITO) thin film
  gas sensor for detection of methanol at room temperature},} Sensors and
  Actuators B: Chemical \textbf{96}, 180--189 (2003).

\bibitem{Yoo2005}
K.~S. Yoo, S.~H. Park, and J.~H. Kang, \enquote{{Nano-grained thin-film indium
  tin oxide gas sensors for H2 detection},} Sensors and Actuators B: Chemical
  \textbf{108}, 159--164 (2005).

\bibitem{Gong2006}
H.~Gong, J.~Hu, J.~Wang, C.~Ong, and F.~Zhu, \enquote{{Nano-crystalline
  Cu-doped ZnO thin film gas sensor for CO},} Sensors and Actuators B: Chemical
  \textbf{115}, 247--251 (2006).

\bibitem{Berreman1972}
D.~W. Berreman, \enquote{{Optics in Stratified and Anisotropic Media:
  4x4-Matrix Formulation},} Journal of the Optical Society of America
  \textbf{62}, 502 (1972).

\bibitem{Li1988}
Z.-M. Li, B.~T. Sullivan, and R.~R. Parsons, \enquote{{Use of the 4x4 matrix
  method in the optics of multilayer magnetooptic recording media},} Applied
  Optics \textbf{27}, 1334 (1988).

\bibitem{Xu2000}
W.~Xu, L.~T. Wood, and T.~D. Golding, \enquote{{Optical degeneracies in
  anisotropic layered media: Treatment of singularities in a 4x4 matrix
  formalism},} Physical Review B \textbf{61}, 1740--1743 (2000).

\bibitem{Lin-Chung1984}
P.~J. Lin-Chung and S.~Teitler, \enquote{{4x4 Matrix formalisms for optics in
  stratified anisotropic media},} Journal of the Optical Society of America A
  \textbf{1}, 703 (1984).

\bibitem{Yeh1979}
P.~Yeh, \enquote{{Electromagnetic propagation in birefringent layered media},}
  Journal of the Optical Society of America \textbf{69}, 742 (1979).

\bibitem{Zhang2015}
S.~Zhang and F.~Wyrowski, \enquote{{Simulations of general electromagnetic
  fields propagation through optically anisotropic media},} in \enquote{Proc.
  SPIE,} , vol. 9630 D.~G. Smith, F.~Wyrowski, and A.~Erdmann, eds. (2015),
  vol. 9630, p. 96300A.

\bibitem{OBrien2013}
D.~B. O'Brien and A.~M. Massari, \enquote{{Modeling multilayer thin film
  interference effects in interface-specific coherent nonlinear optical
  spectroscopies},} Journal of the Optical Society of America B \textbf{30},
  1503 (2013).

\bibitem{Adachi1999}
S.~Adachi, \enquote{{The Reststrahlen Region},} in \enquote{Optical Properties
  of Crystalline and Amorphous Semiconductors: Materials and Fundamental
  Principles,}  (Springer US, Boston, MA, 1999), pp. 33--61.

\bibitem{Huber2005}
A.~J. Huber, N.~Ocelic, D.~Kazantsev, and R.~Hillenbrand, \enquote{{Near-field
  imaging of mid-infrared surface phonon polariton propagation},} Applied
  Physics Letters \textbf{87}, 081103 (2005).

\bibitem{Falge1973}
H.~J. Falge and A.~Otto, \enquote{{Dispersion of Phonon-Like Surface Polaritons
  on $\alpha$-Quartz Observed by Attenuated Total Reflection},} Physica Status
  Solidi (B) \textbf{56}, 523--534 (1973).

\bibitem{Neuner2009}
B.~Neuner, D.~Korobkin, C.~Fietz, D.~Carole, G.~Ferro, and G.~Shvets,
  \enquote{{Critically coupled surface phonon-polariton excitation in silicon
  carbide.}} Optics letters \textbf{34}, 2667--9 (2009).

\bibitem{Passler2017}
N.~C. Passler, I.~Razdolski, S.~Gewinner, W.~Sch{\"{o}}llkopf, M.~Wolf, and
  A.~Paarmann, \enquote{{Second-Harmonic Generation from Critically Coupled
  Surface Phonon Polaritons},} ACS Photonics \textbf{4}, 1048--1053 (2017).

\bibitem{Dai2014}
S.~Dai, Z.~Fei, Q.~Ma, A.~S. Rodin, M.~Wagner, A.~S. McLeod, M.~K. Liu,
  W.~Gannett, W.~Regan, K.~Watanabe, T.~Taniguchi, M.~Thiemens, G.~Dominguez,
  A.~H.~C. Neto, A.~Zettl, F.~Keilmann, P.~Jarillo-Herrero, M.~M. Fogler, and
  D.~N. Basov, \enquote{{Tunable Phonon Polaritons in Atomically Thin van der
  Waals Crystals of Boron Nitride},} Science \textbf{343}, 1125--1129 (2014).

\bibitem{Caldwell2015a}
J.~D. Caldwell and K.~S. Novoselov, \enquote{{Van der Waals heterostructures:
  Mid-infrared nanophotonics},} Nature Materials \textbf{14}, 364--366 (2015).

\bibitem{Caldwell2016}
J.~D. Caldwell, I.~Vurgaftman, J.~G. Tischler, O.~J. Glembocki, J.~C. Owrutsky,
  and T.~L. Reinecke, \enquote{{Atomic-scale photonic hybrids for mid-infrared
  and terahertz nanophotonics},} Nature Nanotechnology \textbf{11}, 9--15
  (2016).

\bibitem{Woessner2015}
A.~Woessner, M.~B. Lundeberg, Y.~Gao, A.~Principi, P.~Alonso-Gonz{\'{a}}lez,
  M.~Carrega, K.~Watanabe, T.~Taniguchi, G.~Vignale, M.~Polini, J.~Hone,
  R.~Hillenbrand, and F.~H.~L. Koppens, \enquote{{Highly confined low-loss
  plasmons in graphene–boron nitride heterostructures},} Nature Materials
  \textbf{14}, 421--425 (2014).

\bibitem{Taubner2006}
T.~Taubner, D.~Korobkin, Y.~Urzhumov, G.~Shvets, and R.~Hillenbrand,
  \enquote{{Near-field microscopy through a SiC superlens.}} Science (New York,
  N.Y.) \textbf{313}, 1595 (2006).

\bibitem{Li2015}
P.~Li, M.~Lewin, A.~V. Kretinin, J.~D. Caldwell, K.~S. Novoselov, T.~Taniguchi,
  K.~Watanabe, F.~Gaussmann, and T.~Taubner, \enquote{{Hyperbolic
  phonon-polaritons in boron nitride for near-field optical imaging and
  focusing},} Nature Communications \textbf{6}, 7507 (2015).

\bibitem{RodriguesdaSilva2010}
R.~{Rodrigues da Silva}, R.~{Mac{\^{e}}do da Silva}, T.~Dumelow, J.~A.~P.
  da~Costa, S.~B. Honorato, and A.~P. Ayala, \enquote{{Using Phonon Resonances
  as a Route to All-Angle Negative Refraction in the Far-Infrared Region: The
  Case of Crystal Quartz},} Physical Review Letters \textbf{105}, 163903
  (2010).

\bibitem{Yoxall2015}
E.~Yoxall, M.~Schnell, A.~Y. Nikitin, O.~Txoperena, A.~Woessner, M.~B.
  Lundeberg, F.~Casanova, L.~E. Hueso, F.~H.~L. Koppens, and R.~Hillenbrand,
  \enquote{{Direct observation of ultraslow hyperbolic polariton propagation
  with negative phase velocity},} Nature Photonics \textbf{9}, 674--678 (2015).

\bibitem{Geim2013}
A.~K. Geim and I.~V. Grigorieva, \enquote{{Van der Waals heterostructures},}
  Nature \textbf{499}, 419--425 (2013).

\bibitem{Caldwell2014}
J.~D. Caldwell, A.~V. Kretinin, Y.~Chen, V.~Giannini, M.~M. Fogler,
  Y.~Francescato, C.~T. Ellis, J.~G. Tischler, C.~R. Woods, A.~J. Giles,
  M.~Hong, K.~Watanabe, T.~Taniguchi, S.~a. Maier, and K.~S. Novoselov,
  \enquote{{Sub-diffractional volume-confined polaritons in the natural
  hyperbolic material hexagonal boron nitride.}} Nature communications
  \textbf{5}, 5221 (2014).

\bibitem{Jacob2014}
Z.~Jacob, \enquote{{Nanophotonics: Hyperbolic phonon–polaritons},} Nature
  Materials \textbf{13}, 1081--1083 (2014).

\bibitem{Passler2017a}
N.~C. Passler and A.~Paarmann, \enquote{{Generalized 4x4 Matrix algorithm for
  light propagation in anisotropic stratified media (Matlab files)},}
  https://doi.org/10.5281/zenodo.400486  (2017).

\bibitem{Neuner2010}
B.~Neuner, D.~Korobkin, C.~Fietz, D.~Carole, G.~Ferro, and G.~Shvets,
  \enquote{{Midinfrared Index Sensing of pL-Scale Analytes Based on Surface
  Phonon Polaritons in Silicon Carbide},} The Journal of Physical Chemistry C
  \textbf{114}, 7489--7491 (2010).

\bibitem{Novikova2013}
N.~Novikova, V.~Yakovlev, E.~Vinogradov, S.~Ng, Z.~Hassan, and H.~A. Hassan,
  \enquote{{Substrate surface polariton splitting due to thin zinc oxide and
  aluminum nitride films presence},} Applied Surface Science \textbf{267},
  93--96 (2013).

\bibitem{Zheng2016}
G.~Zheng, Y.~Chen, L.~Bu, L.~Xu, and W.~Su, \enquote{{Waveguide-coupled surface
  phonon resonance sensors with super-resolution in the mid-infrared region},}
  Optics Letters \textbf{41}, 1582 (2016).

\bibitem{Yariv1984}
A.~Yariv and P.~Yeh, \emph{{Optical Waves in Crystals: Propagation and Control
  of Laser Radiation}} (1984).

\bibitem{Paarmann2016}
A.~Paarmann, I.~Razdolski, S.~Gewinner, W.~Sch{\"{o}}llkopf, and M.~Wolf,
  \enquote{{Effects of crystal anisotropy on optical phonon resonances in
  midinfrared second harmonic response of SiC},} Physical Review B \textbf{94},
  134312 (2016).

\bibitem{Kasic2000}
A.~Kasic, M.~Schubert, S.~Einfeldt, D.~Hommel, and T.~E. Tiwald,
  \enquote{{Free-carrier and phonon properties of n- and p-type hexagonal GaN
  films measured by infrared ellipsometry},} Physical Review B - Condensed
  Matter and Materials Physics \textbf{62}, 7365--7377 (2000).

\bibitem{Gervais1975}
F.~Gervais and B.~Piriou, \enquote{{Temperature dependence of transverse and
  longitudinal optic modes in the $\alpha$ and $\beta$ phases of quartz},}
  Physical Review B \textbf{11}, 3944--3950 (1975).

\bibitem{Otto1968}
A.~Otto, \enquote{{Excitation of nonradiative surface plasma waves in silver by
  the method of frustrated total reflection},} Zeitschrift f{\"{u}}r Physik
  \textbf{216}, 398--410 (1968).

\bibitem{Bluet1999}
J.~Bluet, K.~Chourou, M.~Anikin, and R.~Madar, \enquote{{Weak phonon modes
  observation using infrared reflectivity for 4H, 6H and 15R polytypes},}
  Materials Science and Engineering: B \textbf{61-62}, 212--216 (1999).

\bibitem{Raether1988}
H.~Raether, \emph{{Surface Plasmons on Smooth and Rough Surfaces and on
  Gratings}} (Springer, 1988).

\bibitem{Berini2009}
P.~Berini, \enquote{{Long-range surface plasmon polaritons},} Advances in
  Optics and Photonics \textbf{1}, 484 (2009).

\bibitem{Gervais1974}
{F Gervais and B Piriou}, \enquote{{Anharmonicity in several-polar-mode
  crystals: adjusting phonon self-energy of LO and TO modes in Al 2 O 3 and TiO
  2 to fit infrared reflectivity},} Journal of Physics C: Solid State Physics
  \textbf{7}, 2374 (1974).

\bibitem{Campione2015}
S.~Campione, I.~Brener, and F.~Marquier, \enquote{{Theory of epsilon-near-zero
  modes in ultrathin films},} Physical Review B \textbf{91}, 121408 (2015).

\bibitem{Razdolski2016b}
I.~Razdolski, D.~Makarov, O.~G. Schmidt, A.~Kirilyuk, T.~Rasing, and V.~V.
  Temnov, \enquote{{Nonlinear Surface Magnetoplasmonics in Kretschmann
  Multilayers},} ACS Photonics \textbf{3}, 179--183 (2016).

\bibitem{Palomba2008}
S.~Palomba and L.~Novotny, \enquote{{Nonlinear Excitation of Surface Plasmon
  Polaritons by Four-Wave Mixing},} Physical Review Letters \textbf{101},
  056802 (2008).

\bibitem{Razdolski2016}
I.~Razdolski, Y.~Chen, A.~J. Giles, S.~Gewinner, W.~Sch{\"{o}}llkopf, M.~Hong,
  M.~Wolf, V.~Giannini, J.~D. Caldwell, S.~A. Maier, and A.~Paarmann,
  \enquote{{Resonant Enhancement of Second-Harmonic Generation in the
  Mid-Infrared Using Localized Surface Phonon Polaritons in Subdiffractional
  Nanostructures},} Nano Letters \textbf{16}, 6954--6959 (2016).

\end{thebibliography}




\end{document}